\documentclass[final]{ws-p8-50x6-00}

\begin{document}
\def\lsi{\raise0.3ex\hbox{$<$\kern-0.75em\raise-1.1ex\hbox{$\sim$}}}
\def\gsi{\raise0.3ex\hbox{$>$\kern-0.75em\raise-1.1ex\hbox{$\sim$}}}
\newcommand{\lsim}{\mathop{\lsi}}
\newcommand{\gsim}{\mathop{\gsi}}

\title{Numerical simulations of\\
electroweak baryogenesis at preheating\footnote{
T\lowercase{alk presented by}
A.~R\lowercase{ajantie at} COSMO2000, 
C\lowercase{heju}, 
K\lowercase{orea}, 4--8 
S\lowercase{eptember}, 2000.}
}

\author{A.~Rajantie}
\address{DAMTP, CMS, Wilberforce Road, Cambridge CB3 0WA, UK}
\author{P.M.~Saffin}
\address{CPT, Durham University,
South Road, Durham DH1 3LE, UK}
\author{E.J.~Copeland}
\address{Centre for Theoretical Physics, University of Sussex,
Brighton BN1 9QH, UK}

\maketitle

\abstracts{
It has recently been suggested that the baryon washout problem of
standard electroweak baryogenesis
could be avoided if inflation ends
at a low enough energy density and a parametric resonance transfers
its energy repidly into the standard model fields.
We present preliminary results of numerical simulations in a 
SU(2)$\times$U(1) gauge-Higgs model in which
this process was studied.
}

\noindent DAMTP-2000-126 

\section{Introduction}
In order to explain the observed baryon asymmetry of the universe,
a theory must satisfy three conditions:\cite{Sakharov:1967dj}
\begin{enumerate}
\item{Baryon number violation,}
\item{C and CP violation and}
\item{Deviation from thermal equilibrium.}
\end{enumerate}
In principle, they all seem to fulfilled in 
the electroweak theory and the standard cosmological
Big Bang scenario.\cite{Kuzmin:1985mm} 
At high temperatures, baryon number is violated by non-perturbative
sphaleron processes, which change the
Chern-Simons number and consequently, due to a quantum anomaly, also
the baryon number. 
The electroweak phase transition makes the system fall out of
equilibrium in a natural way.
The strength of CP violation in the electroweak theory is too small,
but
even that is not a severe problem because the constraints for CP
violation arising from beyond the standard model are fairly weak.

However, the baryon asymmetry generated in the electroweak phase
transition gets easily washed out. Although 
sphaleron processes become less frequent after the phase
transition, they don't stop completely. Instead, their rate
is proportional to $\exp(-M_{\rm sph}/T)$, where 
$M_{\rm sph}$ is proportional to the expectation value $\phi$ of
the Higgs field, and unless $\phi$ is large enough, 
the baryon asymmetry is washed out. 
This can only be avoided if the transition is
strongly enough first order so that the discontinuity of the Higgs
field 
is\cite{Shaposhnikov:1987tw}
$\Delta\phi\gsim T$.

In the minimal standard model, the Higgs mass $m_H$ is the only unknown
parameter, and lattice simulations\cite{Kajantie:1997qd} 
have revealed that, whatever
its value, the transition is not strong enough. In more complicated
models, such as MSSM, there are more unknown parameters and this
freedom makes it possible to satisfy the constraint, but only barely.

In an alternative scenario
proposed recently by two 
groups,\cite{Krauss:1999ng,Garcia-Bellido:1999sv} 
the baryon asymmetry is generated by sphaleron processes during a period of
preheating after inflation.
This requires that inflation ends at an energy scale
that is below the electroweak scale and that 
a large fraction of the energy of the inflaton is transferred
rapidly to the standard model fields by a parametric 
resonance.\cite{kofman94} 
In the resulting non-equilibrium power spectrum, all the
fermionic fields
and the
short-wavelength modes of the bosons
are practically in vacuum, but the
long-wavelength bosonic modes have a high energy density.
The sphaleron rate depends strongly on the temperature of these
long-wavelength modes and is therefore very high, and the
out-of-equilibrium
processes can generate a large baryon asymmetry very quickly. Eventually, the
system equilibrates and the effective temperature decreases by a rate
that is much faster than the decay rate of baryons.
The final temperature $T_{\rm reheat}$ is
determined by the initial energy density and provided that it is low
enough, $T_{\rm reheat}\lsim 0.6 T_c$, 
the sphaleron rate becomes negligible and the baryon washout
is avoided.

Simulations of the dynamics of preheating in an Abelian
gauge-Higgs system\cite{Rajantie:2000fd} has confirmed this
qualitative picture, but they cannot address the issue of baryogenesis
directly.
In this talk, we discuss the electroweak theory with the
full gauge group SU(2)$\times$U(1), and argue that
reliable simulations are possible using reasonable approximations.
We present some preliminary results and
discuss the prospect of determining the generated baryon asymmetry
using these simulations.

\section{Simulations}
Let us consider a simple model of inflation, in which the expansion of
the universe is driven by the potential energy of a scalar field,
the inflaton,
rolling slowly down its potential towards its minimum at the origin.
Inflation dilutes away all inhomogeneities and thus
all the standard model fields are in vacuum when inflation ends, and the
inflaton has a large homogeneous expectation value. It goes on rolling down its
potential and starts to oscillate about the minimum of its
potential. We assume that it is coupled to the Higgs field and the two
fields start to resonate,\cite{kofman94} whereby a large amount of
energy is rapidly transferred from the inflaton to the
long-wavelength modes of the Higgs. The details of this process of
preheating depend on the properties of the inflaton, which are
unfortunately unknown. Therefore we simply assume that the energy
transfer is extremely efficient and results in a state in which the
energy is concentrated in the Higgs modes with very long wavelengths. 
From the point of view of the microscopic physics that we want to
describe, this is practically equivalent to the Higgs having a very
large value $\phi_0$. Furthermore, we assume that the effect of the
inflaton to the later dynamics of the system is negligible, and
therefore we don't include it in our simulations as a dynamical
field. 
Thus we can simply
consider the time evolution of the standard model fields with the
special initial conditions in which the Higgs has initial value
$\phi_0$ and all the other fields are in vacuum.

As previous simulations\cite{Rajantie:2000fd,Prokopec:1997rr} 
have shown, the
system will quickly reach a quasi-equilibrium state in which the
long-wavelength modes of the Higgs and gauge fields are approximately
in equilibrium at a high effective temperature. 
The decays of the bosons transfer energy slowly into
the fermions and gluons, and
the effective temperature of the long-wavelength bosonic modes
decreases. As the fermions and gluons are initially in vacuum, this
process can be described perturbatively by a damping term, whose
magnitude $\Gamma\approx 2$~GeV is obtained from the observed lifetime
of $W$ and $Z$ bosons. This approximation is valid until
$t\sim\Gamma^{-1}$.

Thus, the only fields that we have to consider are the
gauge bosons and the Higgs field. For baryogenesis, the relevant
degrees of freedom are the long-wavelength modes, and they have large
occupation numbers. Therefore they behave classically, and thus we can
study the dynamics of the system simply by solving the classical
equations of motion
\begin{eqnarray}
\label{equ:eom}
\partial^2_0\phi &=& D_iD_i\phi
+2\lambda \left(\frac{1}{2}v^2
-\phi^\dagger\phi\right)\phi-\Gamma\partial_0\phi,
\nonumber\\
\partial_0^2 B_i&=&-\partial_jB_{ij}+g'{\rm Im}\phi^\dagger D_i\phi
-\Gamma\partial_0B_i,
\nonumber\\
\partial_0^2 W_i&=&-[D_j,W_{ij}]+ig
\left(
\phi(D_i\phi)^\dagger
-\frac{1}{2}(D_i\phi)^\dagger\phi-{\rm h.c.}
\right)
-\Gamma\partial_0W_i,
\end{eqnarray}
where $\phi$ is the Higgs field, $B_i$ is the U(1) gauge field
and $W_i$ is the SU(2) gauge field, and the covariant derivative is 
\begin{equation}
D_i=\partial_i-\frac{i}{2}gW_i-\frac{i}{2}g'B_i.
\end{equation}
In addition, both gauge fields must satisfy the corresponding Gauss laws
\begin{eqnarray}
\label{equ:gauss}
\partial_iE_i&=&g'{\rm Im}\pi^\dagger\phi,\nonumber\\
\left[D_i,F_i\right]&=&
ig\left(
\pi\phi^\dagger-\frac{1}{2}\phi^\dagger\pi-{\rm h.c.}
\right),
\end{eqnarray}
where
\begin{equation}
\pi=\partial_0\phi,\quad
E_i=-\partial_0B_i,\quad
F_i=-\partial_0W_i.
\end{equation}

Although the classical equations of motion describe 
the dynamics of the long-wavelength modes, they fail to describe the
early stages of the thermalization, when the quantum fluctuations play
an important role. Therefore we 
approximate them by adding to the initial field
configuration
Gaussian fluctuations with the same 
two-point
correlation function as in the quantum theory at tree level.
For each real field component $Q$ of mass $m$ and its canonical momentum
$P$, this means
\begin{eqnarray}
\label{equ:vaccorr}
\langle Q^*(t,\vec{k})Q(t,\vec{k}')\rangle
&=&
\frac{1}{2\sqrt{\vec{k}^2+m^2}}(2\pi)^3\delta^{(3)}(\vec{k}-\vec{k}'),\\
\langle P^*(t,\vec{k})P(t,\vec{k}')\rangle
&=&
\frac{\sqrt{\vec{k}^2+m^2}}{2}(2\pi)^3\delta^{(3)}(\vec{k}-\vec{k}').
\end{eqnarray}
In a sense, this means that the quantum effects are approximated to
leading order in perturbation theory. Just like real quantum
fluctuations, 
these
fluctuations generate radiative corrections to the couplings, and they
must be taken into account, i.e.,~the parameters must be renormalized.
The situation is made more complicated by the damping term $\Gamma$,
which damps also the fluctuations and therefore the mass divergence
and consequently the mass counterterm also decrease with time.
The parameter with the largest radiative corrections in the Higgs
mass, and we have calculated the necessary renormalization counterterm
at one-loop level in perturbation theory
\begin{equation}
m_{\rm latt}^2\approx m_H^2-\left(6\lambda
+\frac{9}{4}g^2
+\frac{3}{4}g'^2 \right)\frac{0.226}{hx^2}e^{-\Gamma t}.
\end{equation}
In a similar way, when we plot $\langle\phi^\dagger\phi\rangle$, 
the quantity is actually
\begin{equation}
\langle\phi^\dagger\phi\rangle\approx
\langle\phi^\dagger\phi\rangle_{\rm latt}-
\frac{0.452}{hx^2}e^{-\Gamma t},
\end{equation}
where we have subtracted the dominant ultraviolet divergence.

With these approximations, the dynamics of the system depends only on
two unknown parameters, the Higgs mass $m_H$, for which we used the
value $m_H=100$~GeV, and the initial value $\phi_0$, which is constrained by
the requirement that when the system equilibrates, it is already deep
enough in the broken phase to prevent the washout of the baryon
asymmetry by sphalerons. It has been estimated\cite{Shaposhnikov:1987tw} that
this requires $\phi\gsim T_{\rm reheat}$, where $T_{\rm reheat}$ is
the final temperature. On the other hand, conservation of energy
implies
\begin{equation}
T_{\rm reheat}\approx\left(\frac{30\lambda}{g_*\pi^2}\right)^{1/4}\phi_0
\approx 0.2\phi_0,
\end{equation}
which leads to the constraint
\begin{equation}
\label{equ:constraint}
\phi_0\lsim 600~{\rm GeV}.
\end{equation}
Since we were only interested in the qualitative behaviour and not
in precise numbers, we used in the simulations the value
$\phi_0=700$~GeV,
which is slightly larger than the constraint (\ref{equ:constraint}).

\section{Results}
\begin{figure}
\center
\epsfxsize=20pc
\epsfbox{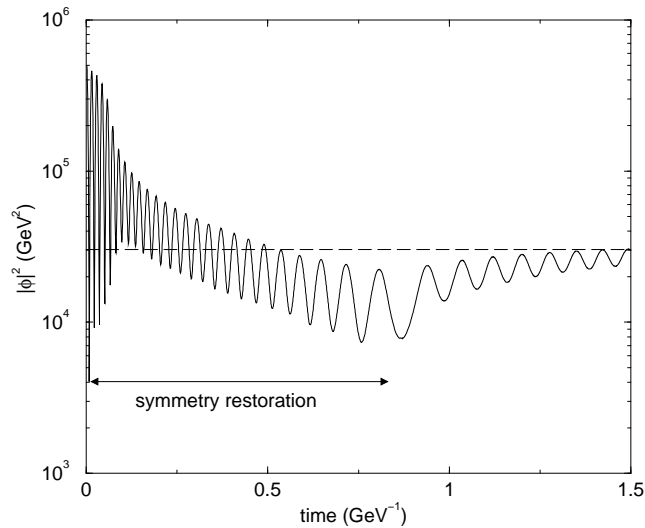}
\flushleft
\caption{
\label{fig:phisq}
The time evolution of $|\phi|^2$ with the initial value
$\phi_0=700$~GeV.
The dashed line shows the vacuum expectation value. Until 
$t\sim 0.8$~GeV$^{-1}$, the curve decreases exponentially, indicating
that the Higgs condensate is absent and the symmetry is restored. 
Eventually, $|\phi|^2$ starts to grow towards it vacuum value, which
means
that the condensate develops and the symmetry is broken.
}
\end{figure}

Startng from the initial configuration described above, we solved
numerically the equations of motion (\ref{equ:eom}) on a $60^3$
lattice with lattice spacing $\delta x=3$~TeV$^{-1}$ and time step
$\delta t=0.6$~TeV$^{-1}$. The time evolution of $|\phi|^2$
is shown in Fig.~\ref{fig:phisq}.
Its
qualitative behaviour is similar to that in the Abelian
theory.\cite{Rajantie:2000fd} Note that in the presence of
fluctuations, $|\phi|^2$ is never zero, but from its qualitative
behaviour we can deduce that the electroweak symmetry is effectively
restored until $t\sim 0.8$~GeV$^{-1}$. In the absence of a Higgs
condensate, the damping term is namely expected to cause $|\phi|^2$ to
decrease as $\exp(-\Gamma t)$, which is exactly what we observe. At
$t\sim 0.7$~GeV$^{-1}$, $|\phi|^2$ starts to grow towards its vacuum
expectation value, just as it is expected to do in the broken phase.
During this period of non-thermal symmetry restoration, baryon number
is not conserved, and the out-of-equilibrium processes can generate a
non-zero baryon asymmetry.

\section{Baryon asymmetry}
So far, we have concentrated on understanding the qualitative dynamics
of the electroweak theory during preheating. However, If we really
want to test the scenario of electroweak baryogenesis at
preheating,\cite{Krauss:1999ng,Garcia-Bellido:1999sv}
we have to be able to measure the baryon asymmetry
generated during the transition, and that involves many technical
problems. 

In principle, we can measure the change of the baryon number even
though we don't have fermions in our system, because a quantum anomaly
links it to the  changes of the Chern-Simons number of the SU(2) gauge
field
\begin{equation}
\Delta B=3\Delta N_{\rm CS}
=
{1 \over 16 \pi^2} \int_0^t dt\int d^3 x 
\epsilon_{ijk} E^a_i F^a_{jk}.
\end{equation}
By measuring $E^a_i$ and $F^a_{jk}$, we could then find the change of
the baryon number. 

However, as $N_{\rm CS}$ is a topological quantity, it does not have a
natural definition on a lattice, and 
attempts\cite{Moore:1997cr,Ambjorn:1997jz} to measure 
$\epsilon_{ijk} E^a_i F^a_{jk}$ in lattice theories have shown that
it is dominated by ultraviolet fluctuations. However, at least in
thermal equilibrium, it is possible to remove these fluctuations by
cooling the system,\cite{Ambjorn:1997jz} which leads to a more reliable result.

Another, more serious problem is that in order to generate the
observed baryon asymmetry, the theory must violate CP. If this effect arises
from heavy degrees of freedom, it can be approximated by an
effective term in the Lagrangian
\begin{equation}
\Delta{\cal L} = {\delta_{\rm CP}\over M_{\rm new}^2}\phi^\dagger\phi \,
{3g^2\over32\pi^2}\,{\rm Tr}F_{\mu\nu}\tilde F^{\mu\nu} \,,
\end{equation}
where $M$ is the mass of the heavy fields and $\delta_{\rm CP}$
parameterizes the strength of the CP violation. Unfortunately, it is
very difficult to add this term in the equations of motion, because it
is extremely sensitive to ultraviolet fluctuations.

Thus, it seems that the only way to measure the generated baryon
asymmetry is to treat $\delta_{\rm CP}$ as a linear perturbation.
One way to do that is to use a Boltzmann-type equation\cite{Khlebnikov:1988sr}
\begin{equation}
\frac{d n_{B}}{dt } = 
\frac{\Gamma_{\rm sph}}{T_{\rm eff}} {\delta_{\rm cp}\over M_{\rm new}^2}
{d\over dt}\langle\phi^2\rangle,
\label{boltzmann}
\end{equation}
and measure $\Gamma_{\rm sph}$, $T_{\rm eff}$ and
$\langle\phi^2\rangle$
in the simulation. However, this approximation may not always take all
the relevant effects into account.\cite{Grigoriev:2000ns}

\section{Conclusions}
We have studied numerically some aspects of
the behaviour of the electroweak theory
during preheating and found that it is possible to restore the
symmetry non-thermally for a short time, which allows the baryon
asymmetry to be generated. When the Higgs and gauge bosons decay into
fermions, the temperature decreases so rapidly that this baryon
asymmetry does not have time to be washed out. 

While the results presented in this talk support the scenario of
electroweak baryogenesis at preheating, they are qualitative in nature and
do not let us deduce the generated amount of baryon asymmetry. 
However, more precise simulations are under 
way.\cite{future}

\section*{Acknowledgments}
The authors were supported by PPARC and AR also partially by the
University of Helsinki.
This work was conducted on the SGI Origin platform using COSMOS
Consortium facilities, funded by HEFCE, PPARC and SGI.

\end{document}